\documentclass[
]{ceurart}

\usepackage{listings}
\usepackage{xurl}

\begin{document}

\copyrightyear{2025}
\copyrightclause{Copyright for this paper by its authors.}

\conference{Meta-HCI '25: First Workshop on Meta-Research in HCI,
  April 26, 2025, Yokohama, Japan}

\title{Research as Resistance: Recognizing and Reconsidering HCI's Role in Technology Hype Cycles}


\author{Zefan Sramek}[
email=zefanS@iis-lab.org,
orcid=0000-0003-4541-0399]

\author{Koji Yatani}[
email=koji@iis-lab.org,
orcid=0000-0003-4192-0420]
\address{IIS Lab, The University of Tokyo, Japan}

\begin{abstract}
  The history of information technology development has been characterized by consecutive waves of boom and bust, as new technologies come to market, fuel surges of investment, and then stabilize towards maturity. However, in recent decades, the acceleration of such technology \textit{hype cycles} has resulted in the prioritization of massive capital generation at the expense of longterm sustainability, resulting in a cascade of negative social, political, and environmental consequences. Despite the negative impacts of this pattern, academic research, and in particular HCI research, is not immune from such hype cycles, often contributing substantial amounts of literature to the discourse surrounding a wave of hype. In this paper, we discuss the relationship between technology and capital, offer a critique of the technology hype cycle using generative AI as an example, and finally suggest an approach and a set of strategies for how we can counteract such cycles through \textit{research as resistance}.
\end{abstract}


\maketitle

\section{The Entanglement of Technology and Capital} \label{entanglement}

\begin{quote}
    A ``technological mystique'' interprets technological failures in the instrumental mode as experimental steps in the inevitable progress of technology, so a critical view that certain dangerous technologies should be restrained is brushed aside by utopian optimism. This mystique assumes that, in the larger picture, continued experimentation will inevitably bring success. But this assumption is flawed, because technology is not completely guided by scientific rational principles. --- Lee Worth Bailey, 2005~\cite{bailey2005enchantments}
\end{quote}

In his book \textit{All That Is Solid Melts Into Air}, Marshall Berman writes that the dark side of capitalism its defenders are afraid to recognize is not the ``drive to exploit people,'' but the fact that everything made by bourgeois society is ``made to be broken tomorrow, smashed or shredded or pulverised or dissolved, so they can be recycled or replaced next week, and the whole process can go on again and again, hopefully forever, in ever more profitable forms''~\cite{berman2010all}. One could be forgiven for thinking that Berman was describing the contemporary landscape of technology development rather than nineteenth century industrialism.

In many ways technological development and capital are tightly intertwined. All stages, from research and development, to production and marketing require capital to function, and thus a system of investment is essential to the establishment of new enterprises within our current market system. This idea, along with the other tenets of the free market, dates to well before the industrial revolution~\cite{grassby1999idea} and is thus nothing new. However, we have witnessed an increasing cascade of negative consequences from the rapid development of information technology over the last several decades, which is not simply driven by the basic systems of investment and free market exchange, which form the basis of a functional, free economy, but instead by the processes of venture capitalism that produce extreme wealth and encourage the prioritization of short-term profit over long-term viability or social responsibility. 

As Thomas Piketty writes in \textit{Capital in the Twenty-First Century}, ``when the rate of return on capital exceeds the rate of growth of output and income, as it did in the nineteenth century and seems quite likely to do again in the twenty-first, capitalism automatically generates arbitrary and unsustainable inequalities that radically undermine the meritocratic values on which democratic societies are based''~\cite{piketty2017capital}. 

Although since at least the industrial revolution technological development has been entangled with this process, there has been an acceleration of this relationship in the last half-century, particularly focused in Silicon Valley. Henton and Held argue that the particular dynamic that has driven the success of Silicon Valley is the process of \textit{creative destruction} embodied in sequential \textit{hype cycles} of emerging technology~\cite{henton2013dynamics}. Following the work of Joseph Schumpeter, they describe creative destruction as the process by which innovations emerge from the destruction of a previous crash. The hype cycle then proceeds through a series of steps; starting from a \textit{technology trigger}, as the technology becomes more visible and valuations increase, the value of the technology trends toward a \textit{peak of inflated expectations}. This leads to a subsequent crash into the \textit{trough of disillusionment}, but businesses that survive are able to refine their products following a \textit{slope of enlightenment} toward a \textit{plateau of productivity}~\cite{henton2013dynamics}. This cyclic process, they argue, coupled with what they term an `innovation habitat', creates the environment for continual long-term growth, as knowledge and capital from previous cycles is accumulated by the actors within the network. 

However, this optimistic view of the hype cycle and the social-technical milieu in which it develops belies a myriad of costs, many of which are externalized by the very actors who stand to profit most from this type of economic development. Beyond the localized effects that lead to increasing regional economic inequality through what Kwon and Sorenson call \textit{Silicon Valley Syndrome}~\cite{kwon2023silicon}, creative destruction and the technology hype cycle incentivizes a fast-moving and potentially reckless approach to development, resulting in technologies that may generate substantial downstream negative consequences, from environmental impacts~\cite{guidi2024environmental} to social and political effects like the spread of misinformation~\cite{wang2019systematic}. 

Furthermore, as the situation approaches that which Piketty describes, in which the ``rate of return on capital exceeds the rate of growth of output and income,''~\cite{piketty2017capital} the economic emphasis shifts towards generating a higher and higher peak as fast as possible, and away from long-term stability, a strategy Reid Hoffman dubs \textit{blitzscaling}~\cite{hoffman2018blitzscaling}. In this way, the situation not only generates ``arbitrary and unsustainable inequalities''~\cite{piketty2017capital} but also de-prioritizes sustainability in general --- economic, social, environmental, or otherwise.

\section{Generative AI, or Technology as Capital-Generating Hype Cycle} \label{gen-ai-hype}

\begin{quote}
    The big tech companies are publicly committed to an extravagant ``AI race'' that they often prioritize above all else. It's completely normal to hear an executive from one of the biggest companies in the world talk about the possibility of a coming singularity, when the AIs will take over.... This is madness. We forget that AI is a story we computer scientists made up to help us get funding once upon a time, back when we depended on grants from government agencies. It was pragmatic theater. But now AI has become fiction that has overtaken its authors. --- Jaron Lanier, 2018~\cite{lanier2018ten}
\end{quote}

The idea that the economics of blitzscaling can generate immense profits while simultaneously driving external harms, and even without resulting in profitable long-term businesses is hardly without precedent. Indeed, numerous Web 2.0 companies have followed this model, typically producing massive capital gains for shareholders and founders while causing cascades of externalities and often posting consistent financial losses. For example, over the last decade, ride-hailing company Uber has seen massive growth in its market valuation~\cite{olsen2017uber} while simultaneously posting nearly unabated losses~\cite{iqbal2024uber} and being implicated in numerous negative externalities from increasing traffic congestion~\cite{tarduno2021congestion, fageda2021measuring} to fueling precarity and dependence among it's workers~\cite{malin2017free,dubal2017drive}, a phenomenon observed across the gig economy~\cite{schor2020dependence}. Music streaming platform Spotify has followed a similar trend of massive growth~\cite{macrotrends2024spotify} accompanied by consistent losses~\cite{macrotrends2023spotify}, and although the streaming model has transformed the music industry, Spotify has consistently decreased their payments to artists year over year~\cite{audium2024spotify}, has been shown to decrease listening diversity among users through its algorithmic recommendations~\cite{anderson2020algorithmic}, and has pushed artists and labels towards algorithm optimization producing what Thomas Hodgson describes as ``creative ambivalence''~\cite{hodgson2021spotify}. At its worst, blitzscaling produces outcomes like that of the coworking space company WeWork; the failure of its original public offering led to founder and former CEO Adam Neumann's resignation in exchange for nearly \$1.7 billion~\cite{farrell2019softbank} and the eventual crash of its pre-IPO valuation of \$47 billion to \$360.9 million four years later~\cite{reuters2023wework}. The blitzscaling model slices the hype cycle described by Henton and Held, eschewing the plateau of sustainable productivity for extreme peaks and troughs, or an amalgamation of the two fueled primarily by continued investment instead of revenue, incentivizing maximal externalization of costs and disavowal of potential harms. 

Unfortunately, companies developing large language models (LLMs) and generative artificial intelligence (AI) seem poised to follow the same accelerated hype cycle trajectory. Currently, we seem to be on the upward slope of the hype cycle, with OpenAI's valuation skyrocketing to \$147 billion in 2024~\cite{pequeno2024openai}. Despite our relatively early stage in the process however, there have already been numerous critiques and harms associated with generative AI development and deployment. The training of the massive models required for generative AI has raised concerns around copyright infringement of data owners~\cite{samuelson2023generative}, environmental impacts of energy use and e-waste production~\cite{kneese2024carbon}, and the outsourcing of traumatic labeling of violent and sexual content to underpaid workers in the Global South~\cite{perrigo2023openai}. Furthermore, the generative AI models produced by this training, such as GPT-4 and it's accompanying ChatGPT, have been described as ``stochastic parrots''~\cite{bender2021dangers} and ``generative adversarial copy machines''~\cite{zeilinger2021generative} that produce \textit{bullshit}\footnote{The use of the word `bullshit' here refers to the philosophical definition outlined by Harry Frankfurt in his book \textit{On Bullshit}~\cite{frankfurt2005bullshit}, meaning to have no concern for truth value.}~\cite{trevisan2024measuring}, contribute to misinformation~\cite{xu2023combating, monteith2024artificial}, and harm critical thinking abilities~\cite{gerlich2025ai,prather2024widening}. 

Potentially even more concerning is the link between investors and founders of AI companies such as OpenAI, like Peter Thiel, with the philosophical positions known as `effective altruism' and `longtermism'~\cite{rooij2023stop,torres2021against}. Simply put, longtermism is a philosophy concerned with the long-term survival of humanity and the corresponding so-called `existential threats' that may impact it. Although this sounds reasonable in and of itself, the issue, as Émile P. Torres points out, comes from what may or may not be considered such an existential threat; ``even if climate change causes island nations to disappear, triggers mass migrations and kills millions of people, it probably isn’t going to compromise our longterm potential over the coming trillions of years''~\cite{torres2021against}. Thinking like this explains claims like those of Jaan Tallinn that AI may be a more existential threat than climate change~\cite{shead2020skype}, the persistence of OpenAI in discussing potential ``catastrophic outcomes''~\cite{openai2023openai}, and the open letter from the Future of Life Institute, a think-tank associated with longtermism~\cite{torres2021against}, calling for a pause on AI development~\cite{foli2023pause}. It also has the dual effect of furthering the hype cycle around generative AI (it is no surprise that many longtermists are also venture capitalists), driving the short-term intensification of capital, and, as Torres argues, of actually harming our long term survival by prioritizing technological development over tangible threats to human life and well-being~\cite{torres2021against}. As Alan F. Blackwell puts it, ``public debates about the distant future dangers of [artificial general intelligence]... seemed to reach a fever pitch at the same time as AI companies and researchers were \textit{creating} serious ethical problems in the present day''~\cite{blackwell2024moral}.

Despite these issues and the the growing number of dissenting voices, there has been an explosion of research exploring LLMs and generative AI at human-computer interaction (HCI) venues in recent years. Academics, particularly those in technology streams such as HCI, are far from insulated from the effects of technology hype cycles. To the contrary, the high-paced publishing and funding environment means that we are embedded in what can be seen as a micro-economy in which academic publications represent a commodified form of intellectual capital that not only signifies expertise and prestige, but also directly influences financial outcomes related to research funding and hiring. In this sense it is no surprise that the rapid increase in publications on generative AI mimics the explosive increase in the valuation of companies like OpenAI. But the connection is not limited to this micro-economic mimesis; the publication hype cycle is itself entangled in the larger technology hype cycle and its associated capital-generating effects. The 2024 CHI conference alone includes hundreds of papers investigating novel applications of generative AI, and this unfortunately often uncritical engagement with the technology adds a large amount of otherwise academically rigourous material to the cyclone of hype-generating discourse. And this phenomenon is of course not limited to AI, but has historically included numerous topics, from XR, to crowdsourcing, to blockchain. And beyond formal academic publications, this effect can be compounded through other channels, such as research agendas, funding proposals, popular articles and media appearances, and conference topics. HCI research is in a particularly critical position in this dynamic because so much of the literature is concerned with developing and exploring novel use cases and applications of emerging technologies. In this way HCI research has the potential to make technology work \textit{for} humans. But the other side of this is that it can just as easily be used as fuel to add legitimacy to the type of hype cycle we have described, and this is something that we, as academics and researchers, need to remain cognizant of. To what extent does our work play into these unsustainable cycles of capital generation, and to what extent can it act as a countervailing force, to question and resist the most harmful outcomes and discourses? To what extent can our collective voice be, as van Rooij puts it, ``a voice of reason''~\cite{rooij2023stop}?

There will likely still be substantial large-scale benefits to come from the development of AI technologies, and certainly HCI and related research has a major role to play in determining and exploring what these benefits may be and how we can best achieve them. That being said, this type of research needs to be conducted with an awareness of the costs associated with the technologies it engages with; how do the proposed benefits weigh against known harms? Researchers are right to avoid engagement with discourses around theoretical dangers such as AI singularity that have no empirical basis, but sand-boxing a technology in order to study it in a vacuum removed from established harms makes it impossible to consider to what extent we want our collective future to be shaped by the influence of such novel and developing technologies.

\section{Research as Resistance} \label{resistance}

\begin{quote}
    Academics should be a voice of reason; uphold values, such as scientific integrity, critical reflection, and public responsibility. Especially in this moment in history, it is vital that we provide our students with the critical thinking skills that will allow them to recognise misleading claims made by tech companies and understand the limits and risks of hyped and harmful technology that is made mainstream at a dazzling speed and on a frightening scale. --- Iris van Rooij, 2023~\cite{rooij2023stop}
\end{quote}

A substantial amount of HCI research seems to be motivated by uncovering how to best adapt technologies for users, given the technologies' inevitable appearance in the world and on the market. And broad statements intended as rhetorical devices to signal the applicability of particular research seem to belie a subtle streak of technological determinism implicit in much HCI literature. However, we must strive to remember that it is simply \textit{not} the case that ``inevitability always wins''~\cite{andriole2024big}.
Despite hyperbolic discourse suggesting the inevitable rise of new technologies, we must remain attuned to the fact all such technologies are developed through the actions of individual and collective actors, and thus we can and do have the opportunity to reconsider, reform, or re-direct. And in this way, we believe research can play a powerful role both in resisting technological determinism, and in resisting technology hype cycles that prioritize capital over sustainable development. 

Unfortunately, academic research itself is embedded in neoliberal logic, and as a result, acts of resistance may be difficult or self-defeating. In the face of securing continued research funding and career development, research publications become commodities in a specialized economy of knowledge capital, which means that academics may stand to benefit in their own ways from technology hype cycles much like early investors. At the same time, the neoliberal system is itself increasingly impervious to resistance. In his book \textit{Non-things}, Byung-Chul Han argues that the neoliberal regime is ``permissive rather than repressive. It does not condemn us to silence. Rather, we are constantly asked to share our opinions, preferences, needs and desires --- even to tell the stories of our lives.... Capitalism culminates in the capitalism of the like. Because it is permissive, it need not fear resistance or revolution''~\cite{han2022nonthings}. In this sense, capitalist systems may subsume criticism or resistance, re-integrating them into the process of acceleration. Criticism or concern, frustrations or worries feed into a cycle of discourse with the technology and its valuation at its center. We saw this in Section~\ref{gen-ai-hype} through the ways that discussions of the dangers of AI singularity and calls for regulation can be used for the antithetical purpose of generating further attention and investment in AI technologies.

That being said, we believe that research and academic discourse, particularly from fields like HCI that are `on the inside,' can still play an important role in at least attempting to resist technologies that primarily drive accumulation of wealth at the cost of enacting substantial harms to society, while also undermining narratives of technological determinism that make it seem as though such developments are inevitable. There are no inevitable technologies and there are no disembodied social or technical forces. The development and dissemination of technologies is the result of the actions of a diverse assemblage of actors, and thus can be shifted, redirected, or reconsidered; ``power, like society, is the final result of a process and not a reservoir, a stock, or a capital that will automatically provide an explanation. Power and domination have to be produced, made up, composed''~\cite{latour2007reassembling}.

HCI researchers are uniquely poised to act as a bridge between research and practice in machine learning, AI development, and other technical domains on the one hand, and social sciences, human well-being, and policy development on the other. We have opportunities to translate proposals for ethical AI~\cite{prem2023ethical}, and other policy recommendations from social science into actionable implementations, and at the same time to bring detailed technical know-how to debates on future policy, regulation, and real-world uses of technology. There may indeed be cases in which we should take a stand against certain developments by refusing to engage with them. However, there are many cases in which to do so is to risk allowing uncritical voices to dominate the discourse. Research as resistance thus becomes a critical practice by which the balance of discourse can be shifted. There are numerous other strategies for resistance that we can explore in our research, of which here we offer a series of preliminary suggestions to be further explored.

\textbf{Balancing:} Rather than focusing solely on optimistic outcomes and potential future application scenarios, we can seek to balance our findings with critical inquiry into related negative consequences or externalities. For instance, Lui \textit{et al.}~\cite{liu2024ai} explore how LLM-based agents can be used to foster creativity in research question development, but couch these results in a discussion of biases and blindspots.

\textbf{Interrogation:} We can resist dominant narratives about new technologies by critically interrogating their assumptions and outcomes. This can be done with regard to specific discourses, like Chakrabarty \textit{et al.}'s~\cite{chakrabarty2024art} work on LLMs and creative writing, or with regard to larger systems, like Wolf \textit{et al.}'s~\cite{wolf2022designing} interrogation of capitalism and its impacts on the design of social computing systems.

\textbf{Deconstruction:} Complex technologies like artificial intelligence, particularly those that are made opaque through corporate obfuscation and homogenized into an amorphous concept tied to a specific term such as `AI', can be difficult to understand, critique, and resist, because their inner workings and constituent parts become obscured. We can thus work to deconstruct such monolithic terms, laying the constituent parts bare for analysis and critique, similarly to how Crawford and Joler~\cite{crawford2018anatomy} deconstruct the supply chain of Amazon's Echo, showing how the seemingly disembodied intelligence is actually created through numerous stages of extraction and development, from lithium mining to data labeling.

\textbf{Renaming:} Similarly to deconstruction, when more and more technologies or systems begin to be lumped under a certain term that comes to obfuscate their individual significance, we can resist by simply changing how we refer to them. As Jaron Lanier states, ``AI is a role-playing game for engineers, not in itself an actual technical achievement. Many of the algorithms that are called AI are interesting and actually do things, of course, but they would be better without the AI storytelling''~\cite{lanier2018ten}. Changing the language we use can disrupt the building of hype around a technology, and give us the tools to better communicate not only the technologies' potential flaws, but also their successes.

\textbf{Re-Reading} When exploring the impacts of any new technology, we can seek new perspectives or work to subvert dominant understandings and narratives by re-reading it through critical or alternative lenses. For example, feminist~\cite{bardzell2010feminist} and queer~\cite{light2011hci} theory provides us with powerful tools for critically developing or evaluating new technologies, as in Riggs \textit{et al.}'s~\cite{riggs2024designing} work that brings queer theory to the design of wearables, and Danuta Jędrusiak's~\cite{jkedrusiak2024queering} analysis of artistic projects that seek to queer AI.

\pagebreak

\textbf{Alarm-Sounding:} When evidence of serious risk of harm emerges, our work can sound an alarm, calling for further investigation and both policy and technical changes around a technology or implementation. For example, Precel \textit{et al.}'s~\cite{precel2024canary} analysis of LLM training data suggests American Jews may be disproportionally affected by intellectual property dispossession. They bring this issue to light, offering implications for policy makers, AI developers, as well as those who may be directly affected by this issue, and also encouraging HCI research to continue working towards fairness and accountability in AI technology and beyond.

\textbf{Research Re-Direction:} When we observe that harms are accumulating around a technology, we can also work to redirect research towards under-studied topics or entirely new areas. We can re-direct research within a specific area or topic, like that work of Ma \textit{et al.}~\cite{ma2024evaluating}, which calls on researchers to go beyond technical refinements to LLMs and instead investigate and confront societal baises towards LGBTQ+ people, or we can fully re-direct beyond a technology, like the work of Blackwell, who, ``rather than dwelling on the (very real) injustices being constructed and reinforced through deployment of AI'' re-directs by ``offering concrete suggestions on how we can make software better in other ways''~\cite{blackwell2024moral}.

\textbf{Co-Opting:} The adoption of any technology will involve the adaptation of that technology by its users. By exploring alternative or subversive uses of technology, we can resist the dominant narratives surrounding it and co-opt it for other purposes. For example, Chong \textit{et al.}~\cite{chong2021exploring} explore how automatic gender recognition systems, which have been criticized for perpetuating harms against trans and non-binary people associated with misgendering~\cite{perilo2024facial}, can be co-opted as a tool to instead help transgender people pass as the gender they identify as.

Not all new technologies, and certainly not all aspects of them, are harmful and problematic. Innovations in computing technology have revolutionized industries and resulted in numerous goods for society. As Blackwell reflects, ``our imagination and enthusiasm have resulted in IT systems, personal devices, and social media that have changed the world in many ways. Unfortunately, imagination and enthusiasm, even if well intentioned, can have unintended consequences. As software has come to rule more and more of the world, a disturbing number of our social problems seem to be caused by software, and by the things we software engineers have made''~\cite{blackwell2024moral}. And in these cases we need mechanisms to change course, and, using strategies like those outlined above, \textit{research as resistance} can be one such tool for doing so. We offer this list of strategies as a starting point, but also call on the HCI community to explore how it can be further extended in theory and further applied in practice.

\section{Conclusions}
\begin{quote}
    Instead of reactionary solutionism, let us ask where the technologies are that people really need. Let us reclaim the idea of socially useful production, of technological developments that start from community needs. --- Dan McQuillan, 2023~\cite{mcquillan2023chatgpt}
\end{quote}

From participatory design to accessible computing, HCI research has long played a valuable role in exploring how technological development can be made more socially conscious and equitable. However, it can also become overly focused on novelty, inadvertently playing into unsustainable and hyper-capitalistic hype cycles. Ultimately, we maintain that all technologies should be \textit{for} people. And as such, rather than falling into and contributing to hype cycles, we should ask ourselves how our research can contribute to human good, how it can solve the problems facing our societies, and how it can be used to bring out the very aspects that contribute to our shared humanity. 
There are many lenses through which we can view novel technologies, and not all of these will be prioritized by the dominant discourse.
Thus, in any given context, it is important to ask: who is this technology for, how sustainable is it, who will benefit from it and who may be harmed, and ultimately, do we want it, and do we need it? And if the answer is no, then to \textit{resist}.

\pagebreak

\begin{acknowledgments}
 The authors would like to thank Susanne Bødker and Olav W. Bertelsen for their insightful feedback and discussions on this paper.
\end{acknowledgments}

%%
%% Define the bibliography file to be used
\bibliography{references.bib}

\end{document}